\documentclass[11pt,a4paper]{article}

\usepackage{amsmath}
\usepackage{amssymb}

\usepackage{graphicx}
\usepackage{float}
\usepackage[topadjust=-1pt,captionskip=0pt]{subfig}
\usepackage[space]{cite}

\usepackage{color} 
\usepackage{eucal} 
\usepackage{tabularx}

\usepackage{enumerate}

\newcommand{\me}[1]{eq.~{\eqref{eq:#1}}}

\renewcommand{\v}[1]{\boldsymbol{#1}}

\newcommand{\choosealtotherw}[3]{
\left\{
  \begin{array}{lll}
    #1 & , & \mbox{ if } #2\\
    #3 & , & \mbox{ otherwise}\\
  \end{array}
\right.}

\newcommand{\MD}{\mathcal{M}}

\usepackage{setspace} 
\title{Feedback inhibition shapes emergent computational properties of cortical microcircuit motifs}
\author
{Zeno Jonke$^*$, Robert Legenstein$^{*\dagger}$, Stefan Habenschuss, Wolfgang Maass\\
\\
\normalsize{Institute for Theoretical Computer Science}\\
\normalsize{Graz University of Technology}\\
\normalsize{Inffeldgasse 16b/I}\\
\normalsize{8010 Graz, Austria}\\
\normalsize{\today}\\
}

\date{}

\begin{document}

\maketitle

\vspace{3cm}

\noindent
\begin{abstract}
Cortical microcircuits are very complex networks, but they are composed of a relatively small number of stereotypical motifs. Hence one strategy for throwing light on the computational function of cortical microcircuits is to analyze emergent computational properties of these stereotypical microcircuit motifs. We are addressing here the question how spike-timing dependent plasticity (STDP) shapes the computational properties of one motif that has frequently been studied experimentally: interconnected populations of pyramidal cells and parvalbumin-positive inhibitory cells in layer 2/3.
Experimental studies suggest that these inhibitory neurons exert some form of divisive inhibition on the pyramidal cells. We show that this data-based form of feedback inhibition, which is softer than that of winner-take-all models that are commonly considered in theoretical analyses, contributes to the emergence of an important computational function through STDP: The capability to disentangle superimposed firing patterns in upstream networks, and to represent their information content through a sparse assembly code. 

\vspace{3cm}
\end{abstract}

{\small
\noindent
$^*$ These authors contributed equally to the work.\\
\noindent
$\dagger$ Corresponding author (email: \verb|robert.legenstein@igi.tugraz.at|).
}

\newpage

\section{Introduction}
A promising strategy for understanding the computational function of a cortical column
was proposed by \cite{DouglasETAL:89,Shepherd:04,GrillnerGraybiel:06}, and others: To probe computational properties of prominent network motifs of a cortical column, commonly referred to as microcircuit motifs. We are addressing computational properties of one of the most prominent microcircuit motifs: densely interconnected populations of excitatory and inhibitory neurons. We focus on motifs in layer 2/3, where parvalbumin-positive ($\text{PV}^+$) inhibitory neurons (often characterized as fast-spiking interneurons, in particular basket cells) are interconnected with nearby pyramidal cells with very high connection probability in both directions, see e.g. \cite{PackerYuste:11, FinoETAL:12,avermann2012microcircuits}. 
One usually refers to this type of inhibition as lateral or feedback inhibition. The dynamics of this microcircuit motif has frequently been examined in-vivo \cite{WilsonETAL:12,PetersenCrochet:13,PalaPetersen:15}, and modelled in \cite{avermann2012microcircuits}.
We examine computational properties that emerge in a model $\MD$ for this microcircuit motif under spike-timing dependent plasticity (STDP).

One cannot model this microcircuit motif by the frequently considered winner-take-all (WTA) model, since this model would require that the firing of a single pyramidal cell (the "winner") can suppress firing of other pyramidal cells in the motif. But experimental data show that several pyramidal cells need to fire in order to engage feedback inhibition through $\text{PV}^+$ cells \cite{IsaacsonScanziani:11,avermann2012microcircuits}. 
Divisive inhibition has been proposed as a more realistic mathematical model for this softer type of inhibition \cite{WilsonETAL:12,CarandiniHeeger:12}. Our goal is to understand the impact of this softer type of inhibition on neural codes and computational properties that emerge under STDP. 
There exists a large number of preceding studies of emergent computational properties of WTA-like microcircuit motifs, from	
\cite{RumelhartZisper:85} to \cite{nessler2013bayesian}.
But they were based on the assumption of strong WTA-like lateral inhibition. 

The functional role of inhibition in this microcircuit
motif can be better approximated by a variation of the k-WTA model \cite{Maass:00}, where several ($k$) winners can emerge simultaneously from a competition of pyramidal cells for firing. 
We show that this softer competition leads to the emergence of shared feature selectivity of pyramidal cells, like in the experimental data of  \cite{LeeETAL:12}, where small subsets of pyramidal cells (assemblies), instead of single neurons,  respond to specific input features (see Figure \ref{fig:fig_tilted}).

We also show that an important computational operation, blind source separation \cite{foldiak1990forming}, also referred to as independent component analysis \cite{HyvaerinenETAL:04}, emerges in this microcircuit motif through STDP. 
This operation enables a network to disentangle and sparsely represent superimposed spike inputs that may result from separate sources in the environments or upstream neural networks. 
This modular coding scheme avoids a combinatorial explosion of the number of neurons that are needed to encode
superimposed sources, since they become encoded by superpositions of neural codes (assemblies) for each of the sources, rather than by a separate neural code for every superposition that occurs.
An example 
is given in Figure \ref{fig:fig_bars},\ref{fig:fig_performance}
for the case of arbitrarily superimposed vertical and horizontal bars, a well-known benchmark task for blind source separation \cite{foldiak1990forming}. This distributed coding scheme 
also supports intra-cortical communication and computation based on spike patterns or spike packets as proposed in \cite{LuczakETAL:15}, see Figure \ref{fig:fig_stp}.



\section{Methods}\label{sec:methods}

\subsection{Definition of a data-based microcircuit motif model $\MD$}
We consider in this article a model for interacting populations of pyramidal cells with $\text{PV}^+$ inhibitory neurons on layer 2/3 that is based on data from the Petersen Lab \cite{avermann2012microcircuits} and refer to this specific model as the microcircuit motif model $\MD$.

The microcircuit motif model $\MD$ consists of
two reciprocally connected pools of neurons, an excitatory pool and an inhibitory pool.  Inhibitory network neurons are recurrently connected. Excitatory network neurons receive additional excitatory synaptic input from a pool of $N$ input neurons.
Figure \ref{fig:fig_model}A summarizes the connectivity structure of the model together with connection probabilities.
Connection probabilities have been chosen according to the experimental data described in \cite{avermann2012microcircuits} and listed in Table \ref{tab:conn} together with connection-type specific synaptic parameters. For a connection probability $p$ between two pools, each individual pair of neurons from these two pools is randomly chosen to be connected by a synapse with probability $p$.

\begin{table}
\begin{center}
\begin{tabular}{|l|c|c|c|c|}
\hline
connection & conn.~prob. [\%] & symbol & $w$ [a.u.] & delay [ms]\\ 
\hline
Input$\rightarrow$E & 100 & $w$ & $[0.01,1]$ & $[0, 10]$ \\
E$\rightarrow$I & $57.5$ & $w^{\text{EI}}$ & $13.57$ & $1$ \\
I$\rightarrow$E & $60$ & $w^{\text{IE}}$ & $1.86$ & $1$ \\
I$\rightarrow$I & $55$ & $w^{\text{II}}$ & $13.57$ & $1$ \\
\hline
\end{tabular}
\end{center}
\caption{Neuron-type specific synaptic connection parameters in $\MD$: Connection probability (conn.~prob.), synaptic weight ($w$), and synaptic delay. Weights from inputs to excitatory network neurons are plastic and bounded to the given range. The corresponding delays are uniformly distributed in the given range. See \cite{JonkeETAL:17} for a motivation of synaptic efficacy values from a theoretical perspective.}
\label{tab:conn}
\end{table}

Input neurons emit Poisson spike trains with time-varying rates. We tested several temporal profiles of theses rates in different simulations as described below in the corresponding sections.  
Let $t_i^{(1)}, t_i^{(2)}, \dots$ denote the spike times of input neuron $i$.
The {\em output trace} $\tilde y_i(t)$ of input neuron $i$ is given by the temporal sum of unweighted postsynaptic potentials (PSPs) arising from input neuron $i$:   
\begin{align} \label{eq:output_trace}
	\tilde y_i(t) = \sum_{f} \epsilon(t-t_i^{(f)}),
\end{align}
where $\epsilon$ is the synaptic response kernel, i.e., the shape of the PSP. 
It is given by a double-exponential function
\begin{equation}\label{eq:doubleexp}
\epsilon(s) = \choosealtotherw{c_{\epsilon}\left(e^{-s/\tau_f}-e^{-s/\tau_r}\right)}{0 \le s \le T_{\epsilon}}{0},
\end{equation}
with the rise time constant $\tau_r=1$ ms, a fall time constant $\tau_f=10$ ms and a cut-off after $T_{\epsilon}=50$ ms, see also Figure \ref{fig:fig_model}B. The constant $c_{\epsilon}=1.435$ was chosen to assure a peak value of $1$. All synapses in the network have the same response kernel $\epsilon$.
For given spike times, output traces of excitatory network neurons and inhibitory network neurons are defined analogously and denoted by $\tilde z_m(t)$ and $I_j(t)$ respectively.

The network consists of $M=400$ excitatory neurons, modeled as stochastic spike response model neurons \cite{JolivetETAL:06} that we define in the following. The stochasticity of the model stems from its stochastic spike generation, where spikes are generated according to a Poisson process with a time-varying rate (the instantaneous firing rate of the neuron). The instantaneous firing rate $\rho_m$ of a neuron $m$ depends exponentially on its current membrane potential $u_m$,
\begin{align}\label{eq:neuron_firing_prob}
  \rho_m(t)\ &= \frac{1}{\tau} \exp(\gamma \cdot u_m(t))\;,
\end{align}
where $\tau=10$ ms and $\gamma=2$ are scaling parameters that control the shape of the response function.
After emitting a spike, the neuron enters an absolute refractory period for $10$ ms during which the neuron cannot spike again.
These excitatory neurons project to and receive inputs from a pool of inhibitory neurons.
Thus, the membrane potential of excitatory neuron $m$ is given by the sum of external inputs, inhibition from inhibitory neurons, and its excitability $\alpha$
\begin{align}
u_m (t) &= \sum_i w_{im} \tilde y_i(t) -\sum_{j \in \mathcal{I}_m} w^{\text{IE}} I_j(t) + \alpha,\label{eq:membrane}
\end{align}
where $\mathcal{I}_m$ denotes the set of indices of inhibitory neurons that project to neuron $m$, and $w^{\text{IE}}$ denotes the weight of these inhibitory synapses. $I_j(t)$ and $\tilde y_i(t)$ denote synaptic input (output traces) from inhibitory neurons and input neurons respectively, see above. We used $\alpha=-5.57$
(These parameter values can be motivated from a theoretical perspective, see \cite{JonkeETAL:17}).

Apart from excitatory neurons there are $M_{\text{inh}}=100$ inhibitory neurons in the network. While \cite{JolivetETAL:06} provides a stochastic model for pyramidal cells, no such model is available for $\text{PV}^+$ inhibitory neurons. Experimental data indicates that in these neurons, the relationship between the synaptic drive and the firing rate, i.e., the frequency-current (f-I) curve, is rather linear over a large range of input strengths \cite{ferguson2013experimentally,ho2012inhibitory}.  
We therefore modeled inhibitory neurons as stochastic spike response neurons with an instantaneous firing rate given by 
\begin{align}\label{eq:inh_firing_prob}
  \rho_m^{\text{inh}}(t)\ &= \sigma_\text{rect}(u_m^{\text{inh}}(t)),
\end{align}
where $\sigma_\text{rect}$ denotes the linear rectifying function $\sigma_\text{rect}(u)=u$ for $u\ge 0$ and $0$ otherwise. 
The absolute refractory period of inhibitory neurons in the model is $3$ ms. Inhibitory neurons receive excitatory inputs from excitatory network neurons as well as connections from other inhibitory neurons. The membrane potentials of inhibitory neurons are thus given by
\begin{align}
u_m^{\text{inh}} (t) = \sum_{i \in \mathcal{E}_m} w^{\text{EI}} \tilde z_i(t) -\sum_{j \in \mathcal{II}_m} w^{\text{II}} I_j(t) + u_\text{opt},\label{eq:inh_membrane} 
\end{align}
where $\tilde z_i(t)$ denotes synaptic input (output trace) from excitatory network neuron $i$,
$\mathcal{E}_m$ ($\mathcal{II}_m$) denotes the set of indices of excitatory (inhibitory) neurons that project to inhibitory neuron $m$,
$w^{\text{EI}}$ ($w^{\text{II}}$) denotes the excitatory (inhibitory) weight to inhibitory neurons, and $u_\text{opt}$ denotes an external optogenetic activation of inhibitory neurons. $u_\text{opt}$ was set to $0$ in all simulations except for the simulation shown in Figure \ref{fig:fig_model}D, where optogenetic activation was modeled by setting $u_\text{opt}=50$ (arbitrary units). The synaptic weights from excitatory network neurons to inhibitory neurons imply that a single spike in the excitatory pool induces a spike in a given post-synaptic inhibitory neuron with a probability of $0.17$, consistent with experimental findings that several excitatory neurons have to be active in order to induce robust spiking in $\text{PV}^+$ interneurons \cite{avermann2012microcircuits}.

Synaptic connections from input neurons to excitatory network neurons are subject to STDP. A standard version of STDP is employed with an exponential weight dependency for potentiation \cite{habenschuss2013emergence}, see Figure \ref{fig:fig_model}C. All input weights $w_{ij}$ are updated as follows. For each postsynaptic spike at time $t_\text{post}$, all presynaptic spikes in the preceding $100$ ms are considered. For each such pre-before-post spike pair with time difference $t_\text{post}-t_\text{pre}$, the weight is increased by
\begin{equation}
	\Delta w_{ij}(t_\text{post}-t_\text{pre}) = \eta e^{-w_{ij}+1} e^{-\frac{t_\text{post}-t_\text{pre}}{\tau_+}},
\end{equation}
with $\tau_+=10$ ms. The learning rate $\eta$ is $0.01$ except for Figure \ref{fig:fig_bars} where $\eta=0.02$ to speed up learning.
For each presynaptic spike at time $t$, all postsynaptic spikes in the preceding $100$ ms are considered. For each such post-before-pre spike pair with time difference $t_\text{pre}-t_\text{post}$, the weight change is given by
\begin{equation}
	\Delta w_{ij}(t_\text{pre}-t_\text{post}) = -\eta e^{-\frac{t_\text{pre}-t_\text{post}}{\tau_-}},
\end{equation}
with $\tau_-=25$ ms.
Synaptic weights are clipped to $w_{\text{min}}=0.01$ and $w_{\text{max}}=1$. In all simulations, initial input weights were drawn from a uniform distribution in the interval $[w_{\text{min}}, w_{\text{max}}]$.
 This concludes the definition of the microcircuit motif model $\MD$.

\subsection{Details to computer simulations of the model $\MD$}\label{sec:EI_sim}

Here, we provide details to the computer simulations reported in {\em Results}. It is recommended that the reader skips this section at first reading. References to the individual subsections are given at the appropriate places in {\em Results}.

All simulations were performed in PCSIM, a spiking neural network simulator written in C++ that provides a Python interface, which was extended in order to support simulation of the model. All simulations were performed with a discretization time step $\Delta t$ of $1$ ms. 

\subsubsection*{Details to simulations for Figure \ref{fig:fig_model}}

For Figure \ref{fig:fig_model}D, the input to $\MD$ was given by simulated visual bars stimuli at various orientations (see {\em Details to simulations for Figure \ref{fig:fig_tilted}} below for details). Orientation-tuned neurons emerged in a learning phase that lasted $400$ s of simulated biological time. The tuning curve of one excitatory neuron was evaluated in the original circuit. Then, optogenetic stimulation of inhibitory neurons was mimicked by setting the external activation $u_\text{opt}$ in \me{inh_membrane} to $u_\text{opt}=50$ (arbitrary units) in all inhibitory neurons, with the effect of increasing the total rate of inhibition. The tuning curve of the same excitatory neuron was then evaluated in this modified network. All procedures in the learning and evaluation phase were the same as described below in {\em Details to simulations for Figure \ref{fig:fig_tilted}}.

\subsubsection*{Details to simulations for Figure \ref{fig:fig_tilted}}

Here, we tested the behavior of $\MD$ on an input distribution that mimics visual bar patterns of various orientations.
In this simulation, network inputs were generated from $180$ two-dimensional binary pixel arrays of size $20 \times 20$. A prototypical horizontal bar of width $2$ pixels centered on the array was rotated in steps of $1$ degree in order to obtain $180$ pixel arrays that span the space of possible bar orientations. These pixel arrays were then transformed into $400$-dimensional rate vectors where each entry had a rate of $75$ Hz if the corresponding pixel was on (i.e., the bar covered that pixel) and $1$ Hz otherwise.
During a simulation, a rate vector was chosen randomly (uniformly out of the 180 vectors). 
The $i^{th}$ component of this rate vector then defined the firing rate of input neuron $i$ to the network.
One rate vector was presented to the network for $50$ ms. During this time, input neurons produced Poisson spike trains with the rate as defined in the corresponding entry of the chosen rate vector. Between the presentation of two consecutive bar patterns, all input neurons spiked with a rate of $2$ Hz for a duration drawn from a geometric distribution with a mean of $50$ simulation time steps $\Delta t$, corresponding to $50$ ms simulated biological time. 

During the learning phase, the network was presented with such patterns for $400$ s. In a testing phase, STDP in the network was disabled and input patterns were presented to the network in the same manner as in the training phase for $100$ h of simulated time.  Average firing rates of excitatory and inhibitory neurons were computed conditioned on specific bar orientations for Figure \ref{fig:fig_tilted}E-G. 

In the simulation for panel H, the same network input was presented to a WTA network model proposed in \cite{nessler2013bayesian}. This model was termed spike-based expectation maximization (SEM) network. We used a model consisting of $400$  neurons that competed in a WTA-like manner (see \cite{nessler2013bayesian} for details on the model). The SEM network was simulated with a time step of $1$ ms with rectangular PSPs of length $10$ ms, a total output rate of $100$ Hz, initial weights chosen from a uniform distribution in $[-0.5, 0.5]$, non-adaptive biases of $0$, and a learning rate of $\eta=0.02$ (see \cite{habenschuss2013emergence} for details on the simulated SEM model).
The learning phase and the testing phase were performed in the same manner as for the model $\MD$.  

\subsubsection*{Details to simulations for Figure \ref{fig:fig_stp} }

Here we tested our microcircuit motif model $\MD$ on input that was created by the nonlinear superposition of $150$ ms long spatio-temporal patterns.

\vspace{0.2cm}
\noindent
{\bf Creation of basic rate patterns:}
Input spike trains to the circuit were created by the superposition of two basic rate patterns. We first describe the creation of basic patterns, the superposition of these patterns will be discussed below.

Let $\v R^i$ denote the $i^{\text{th}}$ basic rate pattern. Formally, a rate pattern $\v R^i$ is a matrix $\v R^i = [r^i_{n,s}]_{n=1,\dots,N; s=1,\dots,S}$ with $r^i_{n,s}$ denoting the firing rate of the pattern in channel $n$ at frame $s$ and $S$ is the number of frames of the pattern. Each frame defines the firing rates of channels (corresponding to the rates of input neurons) for a discrete time bin of length $\Delta t=1$ ms.

For Figure \ref{fig:fig_stp}, we defined a set of two basic rate patterns ${\v R^1, \v R^2}$, each consisting of $N=200$ channels with $S=150$ frames (i.e., the length of basic patterns was $150$ ms).
The firing rate for each channel was obtained by an Ornstein-Uhlenbeck (OU) process drawn independently for each pattern and each channel. More precisely, it was calculated as $r^i_{j,s} = 1.5 \exp (x^i_{j,s \Delta t})$, where $x^i_{j,t}$ is given by a maximum-bounded OU process. The maximum-bounded OU process for a variable $x_t$ is defined as $dx_t = \Theta_{OU}(\mu_{OU} - x_t)dt+\sigma_{OU} dW_t$ if $x_t<\log(50)$ and $dx_t = 0$ otherwise. Here, $t$ denotes continuous time, $\Theta_{OU}>0$ is the changing rate (speed), $\mu_{OU}>0$ is the mean, $\sigma_{OU}>0$ is the noise variance and $W_t$ is the standard Wiener process. 
The parameters were $\mu_{OU} = 0$, $\Theta_{OU} = 5$, and $\sigma_{OU} = 0.5$.
The initial values for $x_t$ in the OU process were drawn from a normal distribution with zero mean and unit variance. The first $50$ ms of the OU process were discarded.

\vspace{0.2cm}
\noindent
{\bf Superposition of basic rate patterns:}
Input spike trains were created by the superposition of a number of patterns, or more precisely their rates, from the set of basic patterns $P=\{\v R^1, \v R^2, \dots\}$. Since the procedure will below also be used for the superposition of bar patterns, we describe it here for an arbitrary number of basic patterns. 

We first describe the procedure that determines which basic patterns to be superimposed at which times (that is, the timing of bars in Figure \ref{fig:fig_stp}B top). Given is a set of basic patterns $P=\{\v R^1, \v R^2, \dots \}$, each pattern of length $S$ time steps. Let $n_{\text{max}}$ denote the maximum number of basic patterns that can be superimposed at any time $t$. We define $n_{\text{max}}$ registers $v_1, \dots, v_{n_{\text{max}}}$. Each register holds at any time step $t$ either no pattern (empty register) or one basic pattern, with the constraint that the registers hold different patterns at any given time step $t$. The following procedure ensures that at any time, the probability that a given register holds some pattern is $p_{\text{loaded}}$.
At time step $t$, each empty register $v_i$ is loaded with some pattern independently from other time steps and other registers with probability $\frac{1}{1+S (1-p_{\text{loaded}})/p_{\text{loaded}}}$. If a register is loaded at time step $t$, the basic pattern to be loaded to this register is chosen uniformly from the set of basic patterns that are currently not held by any register. This basic pattern is then kept in the register for the length of its duration $S$ (afterwards the register is empty, but can be loaded again right away). Note that whether a basic pattern is in register $v_i$ or $v_j$ at some time $t$ is irrelevant with respect to the produced superimposed patterns.

This defines for any time step $t$, which basic patterns are to be superimposed and also the frame at which each of these patterns is at that time. Superposition of basic patterns is then accomplished as described above to obtain the rate for each input neuron. Poisson spike trains are drawn from the resulting rates.

Patterns were first superimposed linearly, then a nonlinearity was applied.
Consider a time $t$ when a set of patterns should be superimposed (these patterns overlap at this time point). For the linear superposition, the rate of a particular channel in the superposition at time $t$ is given by the sum of the rates of this channel in all patterns that overlap at time $t$. More formally, let $\mathcal{S}(t)$ denote the set of indices of patterns that overlap at time $t$ and let $s_i(t)$ denote the frame at which pattern $i$ is at time $t$ (if the pattern presentation started at time $t'$, the pattern is in frame $s_i(t)=\left\lfloor \frac{t-t'}{\Delta t} \right\rfloor +1$ at time $t$, with $\Delta t$ being the discretization time step). Then the linearly superimposed rate $r^\text{linear}_{j}(t)$ for channel $j$ is given by
\begin{equation*}
	r^\text{linear}_{j}(t) = \sum_{i \in \mathcal{S}(t)} r^i_{j,s_i(t)}.  
\end{equation*}
In the final nonlinear step, the firing rate $r_{j}(t)$ of input neuron $j$ at time $t$ is squashed by a sigmoidal nonlinearity
\begin{equation*}
      r_{j}(t) = \frac{f_\text{H}}{1+\exp\left(-\frac{2 \kappa}{f_\text{H}}(r^\text{linear}_{j}(t)-0.5 f_\text{H})\right)},
\end{equation*}
 where $f_\text{H}=75$ Hz is the maximum attainable rate and $\kappa=5$ sets the width of the sigmoidal function. In order to avoid completely silent periods in the input between pattern presentations, the rate of each input neuron is set to $2$ Hz at times $t$ when no patterns are superimposed (i.e., $\mathcal{S}(t)=\{\}$).

For Figure \ref{fig:fig_stp}, we used $2$ basic patterns with a maximum number of superimposed basic patterns of $n_{\text{max}}=2$ and each register had a load probability of $p_{\text{loaded}}=0.5$ (i.e., each register was loaded with some basic pattern half of the time).

\vspace{0.2cm}
\noindent
{\bf Pattern-selectivity and activity plots in Figure \ref{fig:fig_stp}B, C:} Network activity was analyzed after a learning period of $400$ s of simulated biological time. Neurons were classified as preferring pattern 1 (green pattern), as preferring pattern 2 (blue pattern), or as non-selective based on a procedure similar to the one used in \cite{HarveyETAL:12}:
First, an activity trace for each neuron was obtained by convolving its spike response with a double exponential kernel \me{doubleexp} with $\tau_r=1$ ms, $\tau_f=20$ ms, and cut-off time $T_\epsilon=200$ ms. A neuron was considered to be active if it had at least 2 spikes during the simulation time. We classified a neuron as pattern modulated if it was an active neuron and if it had a twice as high average activity trace during presentations of patterns than during the times without patterns. From the pattern modulated neurons, a neuron was classified as pattern selective if it had significantly different activity traces during presentation of blue and green patterns. This was determined by a two-tailed t-test with significance value set at $p<0.05$. If a neuron was pattern selective, its pattern preference was decided based on the average activity trace during blue and green pattern presentations: the preferred pattern was defined as the pattern for which the mean of the activity trace is higher. Finally, we call a neuron that is not pattern selective a non-selective neuron.

For average activity plots in panel C, the green and blue patterns were presented to the network in isolation, $200$ presentations per pattern. Activity traces of all pattern selective neurons were averaged over all presentations of the pattern and subsequently normalized to their peak average activity. Neurons were then sorted by the time of their peak average activity at presentation of their preferred pattern and average activity was plotted in the sorted order for both patterns. In panel B, spike trains were also plotted separately for green pattern preferring neurons, blue pattern preferring neurons and non-selective neurons. The sorting of the former two groups was the same as in panel C. 

\subsubsection*{Details to simulations for Figure \ref{fig:fig_bars} }

For this simulation, basic rate patterns were superimposed as described above for Figure \ref{fig:fig_stp}. The experiment however differed in the number and choice of basic patterns. Network responses, precision measures, and synaptic weight vectors were evaluated and plotted after a learning period of $400$ s simulated biological time. 

\vspace{0.2cm}
\noindent
{\bf Creation of basic rate patterns:}
Basic patterns consisted of $64$ channels that were representing horizontal and vertical bars in a two-dimensional pixel array of size $8 \times 8$ pixels.
The pattern length was $50$ ms ($50$ frames) and in contrast to the basic patterns for Figure \ref{fig:fig_stp}, the rate in each individual channel was constant over the period of the pattern, i.e., $r^i_{n,s} = r^i_n$ for $s=1,\dots, S$. Each of the 64 channels, $r^i_n$ in a basic pattern $\v R^i$ corresponded to one pixel in an $8 \times 8$ pixel array. We defined $16$ basic patterns in total, corresponding to all possible horizontal and vertical bars of width $1$ in this pixel array. For a horizontal (vertical) bar, all pixels of a row (column) in the array attained the value $75$ while all other pixels were set to $0$. The channel rates $r^i_n$ were then defined by the values of the corresponding pixels in the array.

\vspace{0.2cm}
\noindent
{\bf Superposition of basic rate patterns:}
Basic rate patterns were superimposed as described above for Figure \ref{fig:fig_stp}. A maximum of $n_\text{max}=3$ basic patterns were allowed to be superimposed at any time with a load probability of $p_\text{loaded}=0.9$ (see {\em Details to simulations for Figure \ref{fig:fig_stp}} above for a definition). In addition to the rates defined by this superposition, a noise rate of $r_{\text{noise}}(t) = 3(3- n_{\text{pat}}(t))$ Hz was added to each channel, where $n_{\text{pat}}(t)$ denotes the number of patterns that are superimposed at time $t$.

\vspace{0.2cm}
\noindent
{\bf Precision measure:}
Our aim was to quantify how well neurons prefer particular basic patterns (i.e., are tuned to particular basic patterns). To measure tuning properties, we computed the precision measure \cite{rijsbergen1974F1} $\text{Precision}_{ij}$ for each pair of excitatory neuron $i$ and basic pattern (bar) $j$. To this end, we say that a neuron $i$ indicates the presence of a pattern whenever the neuron spikes. The precision $\text{Precision}_{ij}$ is then the fraction between the number of times the presence of pattern $j$ is correctly indicated by neuron $i$ divided by the number of times that neuron $i$ indicates that pattern. Hence, the precision measures how well one can predict the presence of a pattern $j$, given that neuron $i$ spiked. Note that the precision measures whether the pattern is present whenever there is a spike, and not whether the neuron spikes whenever the pattern is present. Since many neurons are jointly representing a pattern, the latter question does not make sense on the individual neuron level (it will be quantified later in Figure \ref{fig:fig_performance}C on the population level).

Panel C, shows for each pair of excitatory neuron $i$ and basic pattern (bar) $j$ the precision measure $\text{Precision}_{ij}$. 
Formally, the precision measure is defined according to  \cite{rijsbergen1974F1} as
\begin{align}
\text{Precision}_{ij} = \frac{\text{TP}_{ij}}{\text{TP}_{ij}+\text{FP}_{ij}},
\end{align}
where $\text{TP}_{ij}$ denotes the true positive count and $\text{FP}_{ij}$ denotes the false positive count for that pair.
The true positive count $\text{TP}_{ij}$ is given by the number of times that neuron $i$ spikes while basic pattern $j$ is present in the input. A pattern that starts at time $t$ is defined to be present in the interval $[t, t+S\Delta t+\tau]$. Here, $S\Delta t$ is the length of the pattern and $\tau=10$ ms corrects for PSPs that increase the firing rates of excitatory neurons even after the pattern disappeared.
The false positive count $\text{FP}_{ij}$ denotes the number of times that neuron $i$ spikes when pattern $j$ is not present.

We say that a neuron $i$ prefers basic pattern $j$ if the neuron has maximum precision for pattern $j$ and this precision is larger or equal to $0.8$, and if the second largest precision that neuron $i$ has for any other pattern is lower then $0.7$. A neuron is said to be pattern-selective if it prefers some pattern and non-selective otherwise.
In panels B and C, pattern-selective neurons are shown and sorted according to their preferred basic pattern.

\subsubsection*{Details to simulations for Figure \ref{fig:fig_performance} }

Figure \ref{fig:fig_performance} shows the behavior of $\MD$ over the course of learning in the overlapping bars task (same setup as Figure \ref{fig:fig_bars}). 10 independent simulation runs were performed, each for $1000$ s of simulated biological time.

In Figure \ref{fig:fig_performance}A, a neuron is considered to be recruited if it is pattern selective. In Figure \ref{fig:fig_performance}B a pattern is considered to be represented if at least on neuron prefers that pattern.
Since several excitatory neurons in the circuit can specialize on a given basic pattern, the network performance shown in Figure \ref{fig:fig_performance}C was evaluated over ensembles of neurons, where ensemble $E_i$ is given by the set of neurons that prefer basic pattern $i$. To quantify how well basic pattern $i$ is represented by ensemble $E_i$, we computed the F1 measure \cite{van2004geometry}. The F1 measure is at its maximum value of $1$ if the following holds true: a neuron in the ensemble $E_i$ is active if and only if basic pattern $i$ is present in the input. False positives (i.~e., some neuron in the ensemble is active in the absence of the basic pattern) and false negatives (i.~e., the basic pattern is present in the input, but no neuron of the ensemble is active) reduce the measure, and the minimum possible value of the measure is $0$. Hence, the F1 measure for basic pattern $i$ measures how well this basic pattern is represented by the ensemble $E_i$.  

Formally, we computed the $F1$ measure \cite{van2004geometry} for ensemble $E_i$ defined as 
\begin{align}
\text{F1}_{i} = \frac{2 \text{TP}_{i} } {2\text{TP}_{i}+\text{FN}_{i}+\text{FP}_{i}},
\end{align}
where $\text{FN}_{i}$ denotes the false negative count.
The true positive count $\text{TP}_{i}$ is given by the number of times that basic pattern $i$ is present and detected by ensemble $E_i$, where the pattern active during $[t, t+S\Delta t]$ is detected if some neuron of the ensemble fires at least one spike within $[t, t+S\Delta t+\tau]$.
The false negative count $\text{FN}_{i}$ denotes the number of times when the pattern $i$ is active but there is not a single spike from ensemble $E_i$. To calculate the false positive count $\text{FP}_{i}$, we split the time between two presentations of the pattern (time without pattern $i$) into periods of $[t, t+S\Delta t+\tau]$ (where the last period can be shorter). Then the false positive count $\text{FP}_{i}$ denotes the number of such periods during which there is at least one spike from ensemble $E_i$.
In Figure \ref{fig:fig_performance}C, the mean F1 measure over all 16 basic patterns is plotted, thus indicating how well all the patterns are represented by the network.
For comparison, a SEM network as used for Figure \ref{fig:fig_tilted}H was trained on the same input for $2000$ s simulated time.

\subsubsection*{Details to simulations for Figure \ref{fig:fig_lag} }

Figure \ref{fig:fig_lag} shows analysis regarding the temporal relation between excitation and inhibition in $\MD$ in the experiment of Section \ref{sec:superpositions} (Figure \ref{fig:fig_stp}) after learning. Panel A depicts the mean firing rate of excitatory neurons in the network (blue; mean taken over all excitatory neurons) and the mean firing rate of inhibitory neurons (red; smoothed through a $100$ ms boxcar filter) as well as the scaled mean firing rate of inhibitory neurons (dashed green). The mean firing rate of inhibitory neurons is scaled down by the ratio of the average excitatory and inhibitory firing rates (average taken over the whole simulation time) in order to facilitate comparison.

The lag between excitation and inhibition was quantified in a similar manner as in \cite{OkunLampl:08}.
The cross-correlation function between excitatory and scaled inhibitory firing rate for a duration of $10$ seconds was computed (plotted in panel B). The lag was then given by the offset of the peak in the cross-correlation function (plotted in panel C) from $0$.
In order to evaluate the influence of connections between inhibitory neurons, we performed the same simulations without I-to-I connections and quantified the lag in the same manner (panel C). In order to facilitate a fair comparison, in addition to removing I-to-I connections we also scaled down synaptic weights of I-to-E connections by factor of $0.1155$ to obtain the same excitatory firing rate as in the case with I-to-I connections.

\begin{figure}[t]
\begin{center}
\includegraphics[width=0.8\textwidth]{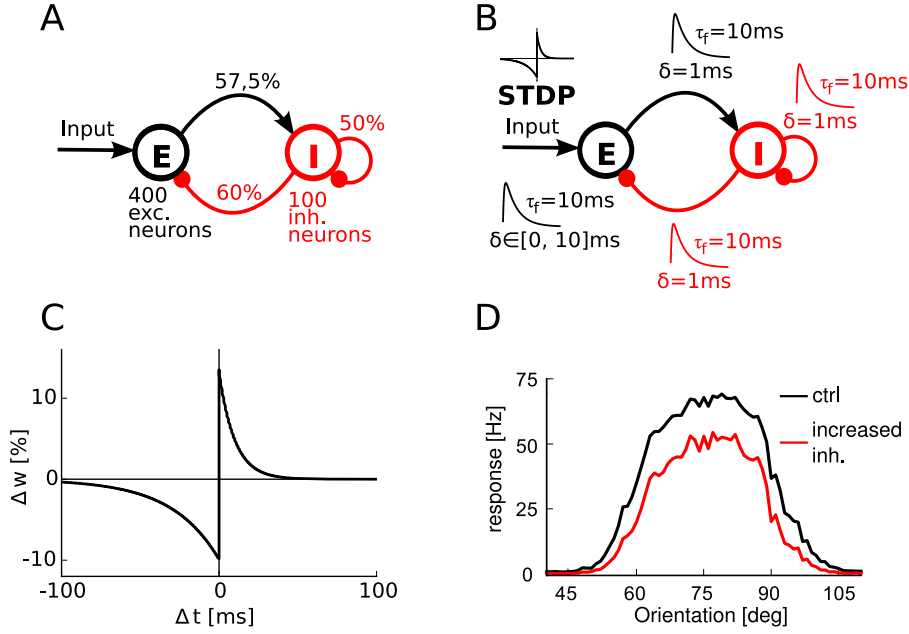}
\end{center}
\caption{{\bf A data-based microcircuit motif model $\MD$.}
\textbf{A}) Network anatomy. Circles denote excitatory (black) and inhibitory (red) pools of neurons. Black arrows indicate excitatory connections. Red lines with dots indicate inhibitory connections. Numbers above connections denote corresponding connection probabilities. \textbf{B}) Network physiology. Same as in (A), but connection delays $\delta$ and PSP shapes with decay time constant $\tau_\text{f}$ are indicated for synaptic connections. Input synapses are subject to STDP. 
\textbf{C}) Standard STDP curve that is used in $\MD$. Shown is the change of the synaptic efficacy in our model for 10 pre-post pairings in dependence on the time-difference $\Delta t=t_{\text{post}}-t_{\text{pre}}$ between a postsynaptic spike at time $t_{\text{post}}$ and a presynaptic spike at time $t_{\text{pre}}$. \textbf{D}) Divisive normalization in the model $\MD$. The response of an excitatory neuron in the circuit to a visual bar-stimulus at various orientations (see {\em Methods} and below for details) in control condition (black) and for simulated increased firing of inhibitory neurons (red). Note the divisive nature of inhibition (stronger responses are more strongly depressed in absolute terms). Compare to Figs.~2e, 3f in \cite{WilsonETAL:12}.
}
\label{fig:fig_model}
\end{figure}

\section{Results}

\subsection{A data-based model for a microcircuit motif consisting of excitatory and inhibitory neurons}\label{sec:ei-model}

We analyze computational properties of densely interconnected populations of excitatory and inhibitory neurons. In particular, we analyze a 
model for interacting populations of pyramidal cells with $\text{PV}^+$ inhibitory neurons on layer 2/3 that is based on data from the Petersen Lab \cite{avermann2012microcircuits}, see Figure \ref{fig:fig_model}A, B. We refer to this specific model as the microcircuit motif model $\MD$.

The excitatory pool in $\MD$ consists of $M$ stochastic spiking neurons, for which we use a stochastic version of the spike response model that has been fitted to experimental data in \cite{JolivetETAL:06}.
In this model the instantaneous firing rate $\rho_m(t)$ of neuron $m$ is approximated by the exponential function applied to the current membrane potential (see eq.~\eqref{eq:membrane} in {\em Methods}).
These excitatory neurons project to and receive inputs from a pool of inhibitory neurons, that are also interconnected among themselves, with connections probabilities taken from \cite{avermann2012microcircuits}.
Each excitatory neuron $m$ in the network also receives excitatory synaptic inputs $\tilde y_1(t),..,\tilde y_N(t)$ from external input neurons, whose contribution to its membrane potential at time $t$ depends on the synaptic efficiency $w_{im}$ between the input neuron $i$ and neuron $m$. We assume that these afferent connections are subject to a standard form of STDP, see Figure \ref{fig:fig_model}C and {\em Definition of a data-based microcircuit motif model $\MD$} in {\em Methods} for details.

Negative (inhibitory) contributions $\sum_{j \in \mathcal{I}_m} w^{\text{IE}} I_j(t)$ to the membrane potential of pyramidal cell $m$ have according to the neuron model a divisive effect on its firing activity, since
its instantaneous firing rate $\rho_m$ can be written (by substituting \me{membrane} in \me{neuron_firing_prob}) as: 
\begin{align}\label{eq:neuron_firing_prob_divisive}
  \rho_m(t) &= 
  \frac{1}{\tau} \frac{\exp\left(\gamma \sum_i w_{im} \tilde y_i(t) + \gamma \alpha \right)}{\exp\left(\gamma \sum_{j \in \mathcal{I}_m} w^{\text{IE}} I_j(t)\right)}\;\;.
\end{align}
Here, the numerator includes all excitatory contributions to the firing rate $\rho_m(t)$, that is, the synaptic inputs (unweighted sum of EPSPs) $\tilde y_i(t)$ from input neurons weighted by the corresponding synaptic weights $w_{im}$. $\alpha$ denotes the neuronal excitability, and $\tau, \gamma$ are scaling parameters that control the shape of the response function of the neuron. The denominator in this term for the firing rate describes inhibitory contributions, thereby reflecting divisive inhibition \cite{CarandiniHeeger:12}. Here, $I_j(t)$ denotes synaptic input from inhibitory neuron $j$ weighted by some common weight $w^{\text{IE}}$ ($\mathcal{I}_m$ denotes the set of all inhibitory neurons that connect to neuron $m$).

Divisive inhibition has been shown to be 
characteristic for the interaction of pyramidal cells with $\text{PV}^+$ inhibitory neurons \cite{WilsonETAL:12}. 
In order to test also on a functional level the divisive character of inhibition in the model, we artificially increased the firing rate of inhibitory neurons in the circuit by a constant, corresponding to the in-vivo experiment described in \cite{WilsonETAL:12}, where activity of the $\text{PV}^+$ neurons was increased through optogenetic stimulation. The response of pyramidal neurons to this increased inhibition  in $\MD$ resembles the experimental data, see Figure \ref{fig:fig_model}D.

\begin{figure}
\begin{center}
\includegraphics[width=0.85\textwidth]{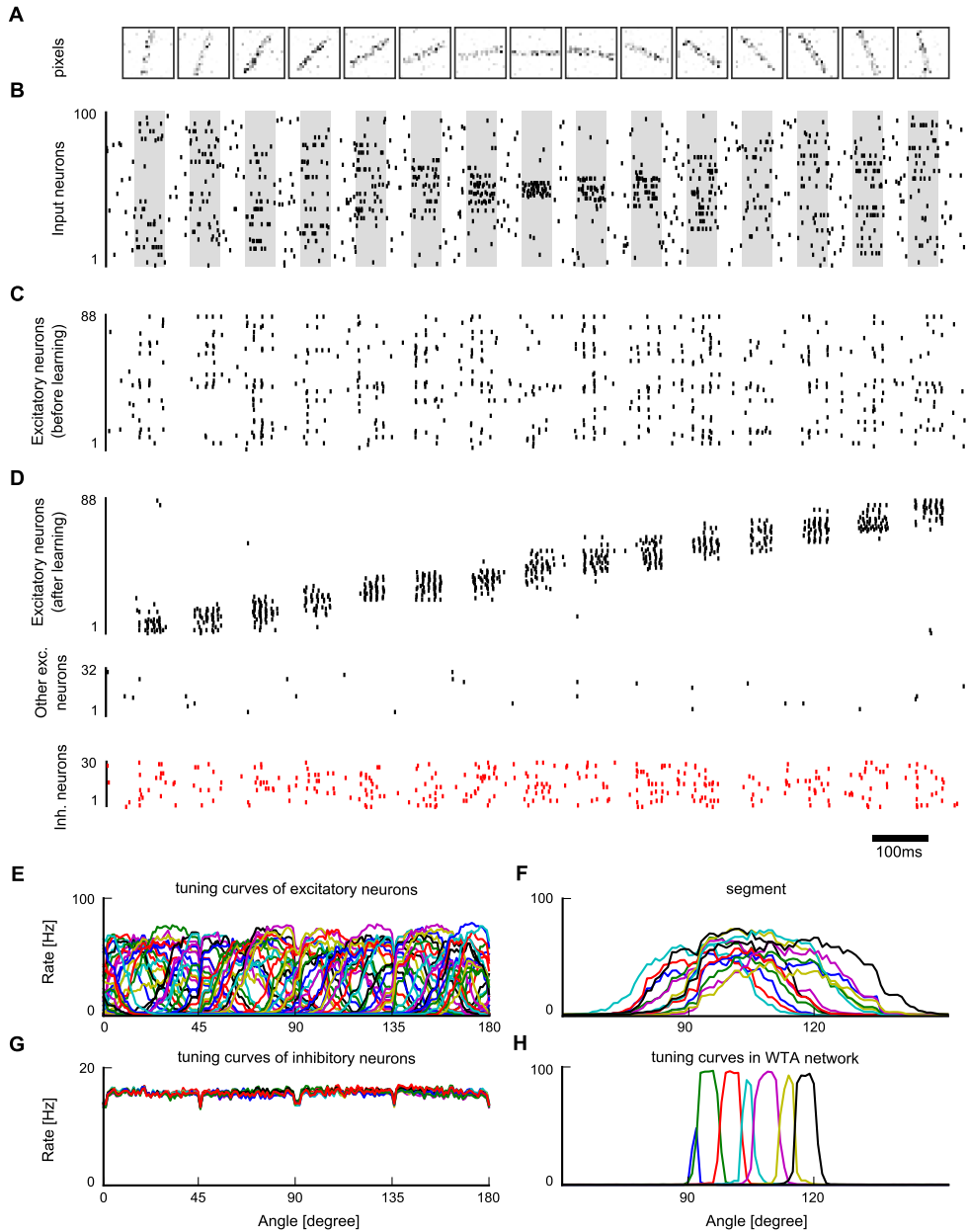}
\end{center}
\caption[]{{\bf Emergent neural codes in the microcircuit motif model $\MD$.} {\bf A}) Bars at various orientations serve as network inputs. Shown are network inputs arranged in 2D for clarity. Gray-level of each pixel indicates the resulting effective network input $\tilde y_i(t)$ (see eq.~(\ref{eq:output_trace})) at some time point $t$. {\bf B}) Resulting spike pattern of input neurons (every $4^{th}$ neuron shown) for different bar orientations. Gray shading indicates the presence of a bar in the input with orientation indicated in panel A. {\bf C}) Example spike pattern of a subset of excitatory neurons in the circuit to this input before learning. {\bf D}) Spiking activity of the same neurons for the same input after applying STDP to all synapses from input neurons to excitatory neurons for $400$ s. Only responses of orientation selective neurons are shown, sorted by preferred orientation. Spiking activity of a random subset of non-orientation selective neurons and inhibitory neurons to the same input is shown below. {\bf E}) Emergent tuning curves of orientation selective excitatory neurons. {\bf F}) The same as in E, but zoomed in on orientations between $90$ and $120$ degrees. {\bf G}) Inhibitory neurons are not orientation selective. {\bf H}) Emergent tuning curves of neurons in a previously considered WTA model \cite{nessler2013bayesian,habenschuss2013emergence}.
}
\label{fig:fig_tilted}
\end{figure}

\subsection{Emergent neural codes: From WTA to k-WTA}\label{sec:emergent-coding-properties}
In our first test of emergent computational properties of this microcircuit motif model $\MD$ we examined the emergence of orientation selectivity. We provided as external spike inputs
pixel-wise representations of bars in numerous random orientations with superimposed noise (Figure \ref{fig:fig_tilted}A). Bars were transformed into high-dimensional spike inputs by representing each black pixel of an oriented bar for $50$ ms through a Poisson input neuron with a Poisson rate of $75$ Hz, whereas all other input neurons had a Poisson rate of $1$ Hz. See Figure \ref{fig:fig_tilted}B for a typical resulting spike input pattern. The initial network response is shown in Figure \ref{fig:fig_tilted}C, and the network response after applying STDP for $400$ s to all synapses from input neurons to excitatory neurons in Figure \ref{fig:fig_tilted}D. 
One clearly sees 
in Figure \ref{fig:fig_tilted}D the emergence of assembly codes for oriented bars. A closer look at the resulting tuning curves of excitatory neurons in Figure \ref{fig:fig_tilted}E, F shows a dense covering of orientations by Gaussian-like tuning curves similar as in experimental data from orientation pinwheels (see Figure 2 d,e in \cite{OhkiETAL:06}). In contrast, inhibitory neurons did not become orientation selective (Figure \ref{fig:fig_tilted}G)
in accordance with experimental data \cite{KerlinETAL:10,IsaacsonScanziani:11}.

The tuning curves of excitatory neurons in Figure \ref{fig:fig_tilted}E, F demonstrate a clear difference between the impact of divisive inhibition in this data-based model $\MD$ and previously considered idealized strong inhibition in WTA-circuits \cite{nessler2013bayesian}, see Figure \ref{fig:fig_tilted}H on emergent computational properties. 
In the data-based model $\MD$ several (on average $k=17$) neurons respond to each orientation with an increased firing rate. This suggests that the emergent computational operation of the layer 2/3 microcircuit motif with divisive inhibition is better described as k-WTA computation, where $k$ winners may emerge simultaneously from the competition. In contrast, for the WTA model with idealized strong inhibition \cite{nessler2013bayesian} at most a single neuron could fire at any moment of time, and as a result at most two neurons responded after a corresponding learning protocol with an increased firing rate to a given orientation (see Figure \ref{fig:fig_tilted}H and Figure 5 in \cite{nessler2013bayesian}).

The k-WTA computation is known to be for $k > 1$ more powerful than the simple WTA computation from the perspective of computational complexity theory \cite{Maass:00}. However the number $k$ of winners is in this microcircuit motif not fixed: It depends on synaptic weights and the external input. Hence one can describe its computation best as an adaptive k-WTA operation.

\subsection{Emergent computation on spike patterns}
\label{sec:superpositions}

\begin{figure}[tbp]
  \begin{center}
   \includegraphics[width=0.9\textwidth]{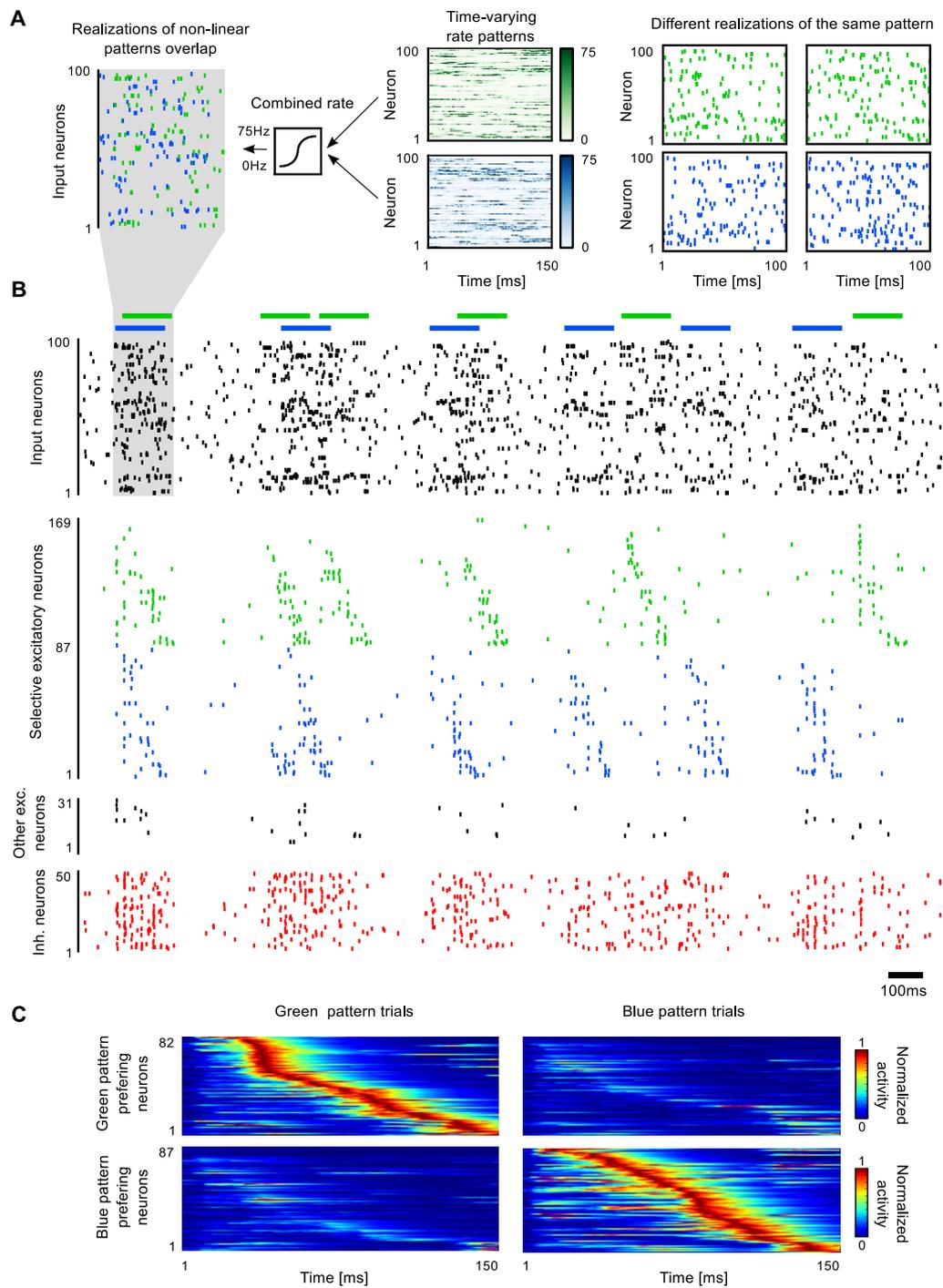}
  \end{center}
\caption{(Figure caption on the next page)}
\label{fig:fig_stp}
\end{figure}

\begin{figure}[tbp]
\ContinuedFloat
\caption[]{{\bf Emergent computation on large-scale spike patterns.}
\textbf{A}) Two spatio-temporal patterns. Each pattern consists of $200$ time-varying firing rates over $150$ ms
generated by an Ornstein-Uhlenbeck process,  
see the middle panel (only every $2^{nd}$ channel shown for clarity). These rate patterns give rise to highly variable spike patterns, as shown 
on the right. Basic rate patterns are superimposed nonlinearly with arbitrary relative timing. The left panel shows one realization of superimposed patterns for a time segment
in panel B. Spikes are colored according to the basic pattern that most probably caused the spike (i.e., the one with the higher rate at that time).
\textbf{B}) Firing response of the neurons in our model $\MD$ for a test input stream after letting STDP be active for synaptic connections from input neurons to excitatory neurons in $\MD$.  
Two subpopulations emerged (green and blue), where each 
neuron specialized on a specific pattern and on a particular time segment within this pattern. Spiking activity of a subset of non-selective neurons (black) and inhibitory neurons (red) are shown below.
\textbf{C}) Average firing rate of neurons preferring the green (top) and blue (bottom) input pattern when the green (left) and blue pattern (right) is shown in isolation. 
Neurons are ordered according to their peak firing rate for the preferred pattern as in panel B.
Resulting selective firing responses are qualitatively similar to data from sensory cortices \cite{LuczakETAL:15} and higher cortical areas \cite{HarveyETAL:12}.
}
\end{figure}
Simultaneous recordings from large numbers of neurons demonstrate the prominence of large-scale activity patterns in networks of neurons \cite{LuczakETAL:15}. They are commonly referred to as assemblies, assembly sequences, or assembly phase sequences. Since \cite{Hebb:49} they have been proposed to reflect tokens of brain computations that connect the fast time scale of spikes (ms) to the slower time scale of cognition and behaviour ($100$'s of ms). But their precise role in neural coding and computation has remained unknown. It is proposed by \cite{LuczakETAL:15} that they serve as basic information components in global cortical communication, where each of these activity patterns is initiated by a particular cortical region and broadcast to all areas it projects to. We show here that our microcircuit motif model $\MD$ is able to carry out a computational operation on large-scale activity patterns that is fundamental for such a global communication scheme: It can demix superimposed spike patterns that impinge on a generic cortical area, and represent the presence of each pattern in their input stream through the firing of separate populations of neurons. This suggests that the layer 2/3 microcircuit motif has an inherent capability to solve the well known cocktail party problem (blind source separation) \cite{Cherry:53} on the level of larger activity patterns. This capability emerges automatically through STDP, as demonstrated in Figure \ref{fig:fig_stp} for our data-based model $\MD$.  

The input to the microcircuit motif model $\MD$ is generated in Figure \ref{fig:fig_stp} by $200$ spiking neurons. Two repeating activity patterns (green and blue patterns) are superimposed for the generation of Poisson spike trains (shown for every $2^{nd}$ neuron in the top row of Figure \ref{fig:fig_stp}B ). These two large-scale activity pattern consist of two time varying rate patterns for the $200$ input neurons (center of Figure \ref{fig:fig_stp}A) that are nonlinearly superimposed with random offsets in the continuous spike input to our model. In spite of these random offsets and the large trial-to-trial variability of spike times in each of the two patterns (see panels on the right of Figure \ref{fig:fig_stp}A), STDP in the synaptic connections from inputs to the excitatory neurons in the model produced after $400$ s two assemblies (green and blue spikes in the middle row of Figure \ref{fig:fig_stp}B). Each responded to just one of the two input patterns, and represented its temporal progress through a stereotypical sequential firing pattern (Figure \ref{fig:fig_stp}C). This effect occurs even if none of the two input patterns is ever presented in isolation during learning, as shown for illustration purposes for test inputs after learning at the right side of Figure \ref{fig:fig_stp}B. Such emergent demixing of superimposed spike patterns in the layer 2/3 microcircuit motif could enable downstream neurons to selectively respond to just one of the patterns. Furthermore the sequential activation of the two assemblies can also inform downstream networks through the firing of specific neurons about the current phase of each of the two input patterns.

\begin{figure}[!t]
\includegraphics[width=\textwidth]{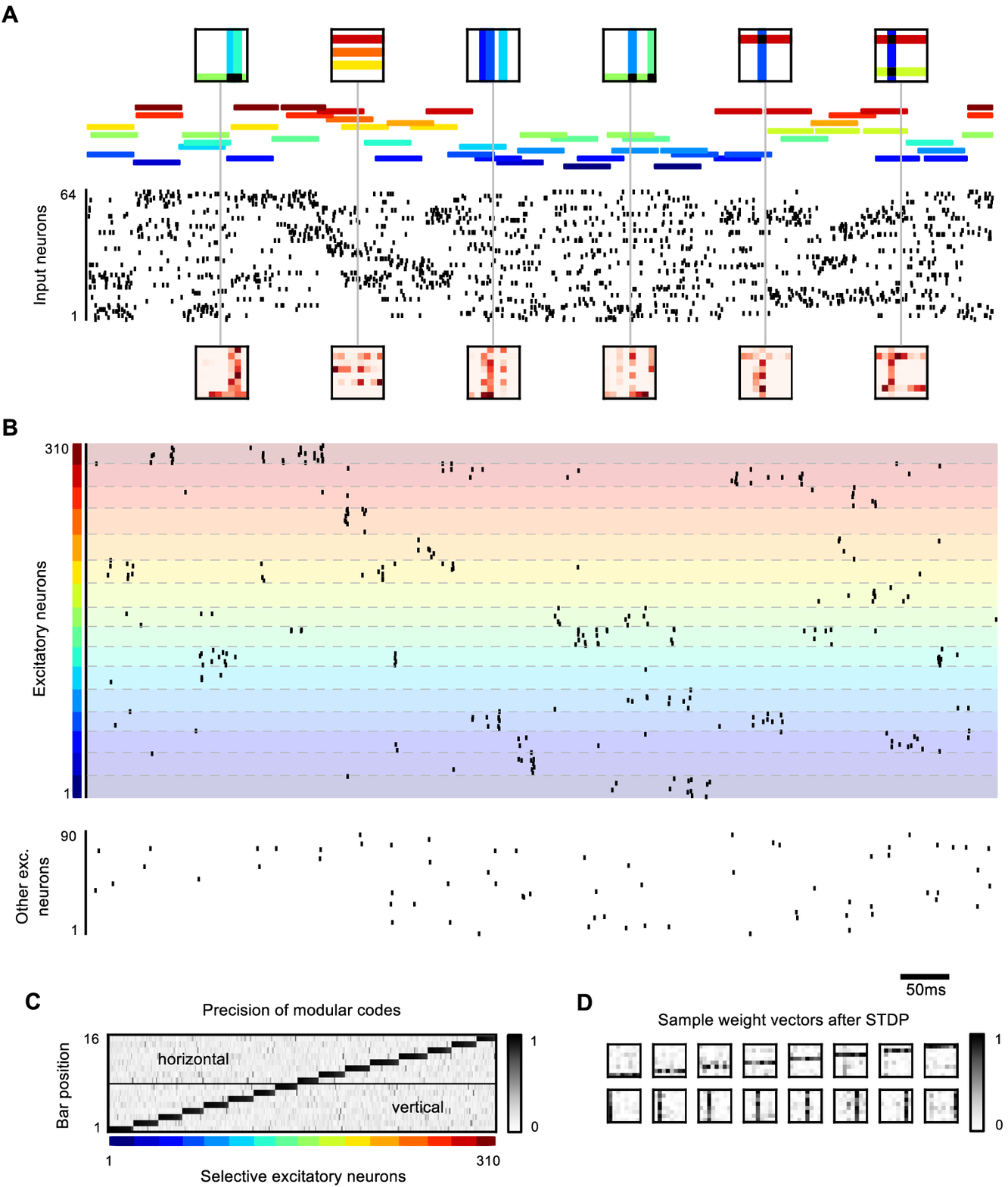}
\caption{(Figure caption on the next page)}
\label{fig:fig_bars}
\end{figure}

\begin{figure}[!htb]
\ContinuedFloat 
\caption[]{{\bf Test of the emergence of modular sparse codes for a common benchmark test.}
\textbf{A}) A difficult version of F\"oldiak's superposition-of-bars problem with asynchronously varying numbers of up to 3 superimposed bars. Each of the 16 bar positions is indicated by a separate color, and its presence in the resulting spike input stream (see middle row) is indicated by a horizontal colored line above the spike raster. 6 of 
696
possible composed input patterns are shown for arbitrarily chosen time points indicated by grey vertical lines. The pattern at the bottom of each line indicates the effective spike input that the network receives at that moment in time, see eq.~\eqref{eq:output_trace}.  
\textbf{B}) Emergent modular assembly codes in the model  $\MD$ after 400s. A small assembly of neurons emerges for each of the 16 bar positions (see color code at left axis and background shading). Activity of non-selective excitatory neurons is shown at the bottom. 
\textbf{C}) Quantitative analysis of the precision of the emergent assembly codes measured according to \cite{rijsbergen1974F1} for each of the 310 neurons from the upper part of B) on the x-axis. Dark shading means high precision for encoding the bar position plotted on the y-axis.
\textbf{D}) Typical weight vectors of neurons from the 16 assemblies that had emerged.
}
\end{figure}

\subsection{Emergent modular sparse coding}

\cite{foldiak1990forming} suggested that complex objects or scenes are encoded in the brain through a sparse modular code, where each neuron signals through its firing the presence of a particular feature in the network input. In this way a combinatorial explosion of the number of neurons is avoided, that would be required if each complex external object or scene is encoded as a whole by separate neurons. \cite{foldiak1990forming} proposed to use superpositions of bars (lines), like in the top part of Figure \ref{fig:fig_bars}A, as benchmark inputs to test sparse modular coding capabilities of neural network models. A neural network is able to avoid the combinatorial explosion of the number of neurons that are needed to encode such complex inputs if it learns to represent them in a modular fashion, where each neuron encodes the presence of one of the bars (in a particular location) in the composed input. A key question is how such codes can emerge in a network autonomously if only composite images (consisting of several superimposed bars) are presented as network inputs. A WTA circuit is not able to develop a good modular code since it does not allow that inputs are represented through the firing of more than one neuron. Hence a natural question is whether biologically more realistic softer lateral inhibition, as implemented in our model $\MD$, supports the emergence of sparse modular codes through STDP. Figure \ref{fig:fig_stp} demonstrated already some weak form of modular coding for superpositions of two spatio-temporal patterns in the input. 

\begin{figure}[!t]
 \includegraphics[width=\textwidth]{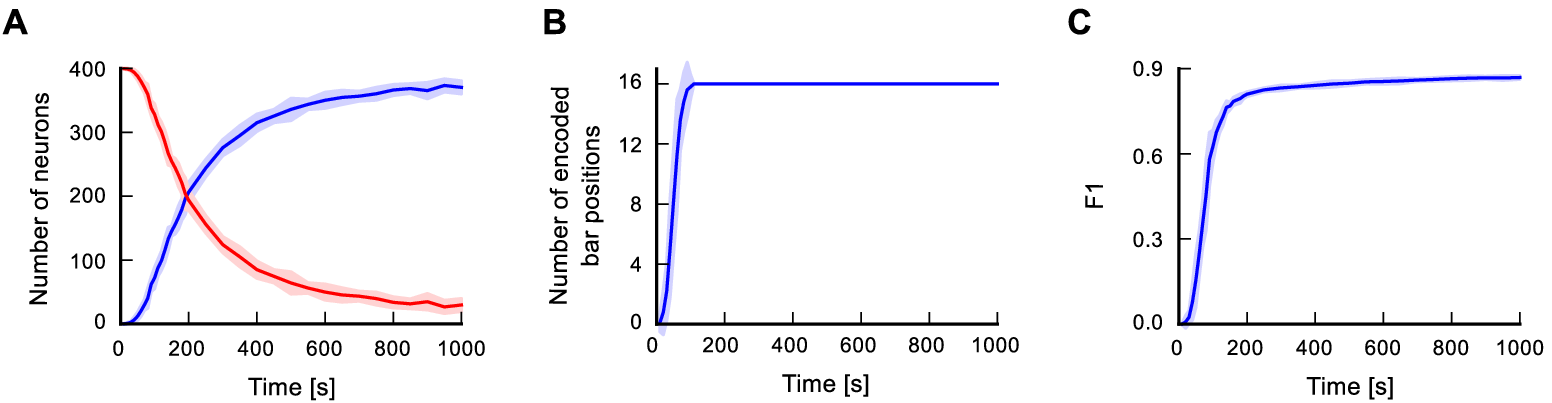}
\caption[]{{\bf Quantitative analysis of emergent coding properties in the benchmark task of Figure \ref{fig:fig_bars}.}
\textbf{A}) Evolution of the number of bar selective (blue) and non-selective (red) neurons during learning. 
 \textbf{B}) The number of bar positions represented by the network rises rapidly during learning. A bar position is considered to be represented if at least one excitatory neuron is selective for it.
\textbf{C}) Average F1 measure of pattern-selective neural ensembles during learning. High F1 measure (maximum is $1$) indicates emergence of highly selective assemblies of neurons for all bar positions (see {\em Methods}). 
In all plots, saturated colors indicate mean and light colored shading indicates STD over $10$ runs.
}
\label{fig:fig_performance}
\end{figure}

Emergent neural codes for F\"oldiak's superposition-of-bars problem are examined in Figure \ref{fig:fig_bars} and Figure \ref{fig:fig_performance} for our model $\MD$ with 400 excitatory and 100 inhibitory neurons as before. Superpositions of up to 3 bars were presented through 64 spiking input neurons in a pixel-wise encoding. Each Poisson input neuron signaled for 50 ms through an increased firing rate if the corresponding pixel was covered by a bar (each bar covered 8 horizontal or 8 vertical pixels in an 8$\times$8 pixel array). Each of the 16 possible bar positions is indicated in  Figure \ref{fig:fig_bars}A through a different color. 
Composed network inputs were created by randomly drawing superposition of bars from the pool of 
696
combinations of up to 3 bars. Obviously our model $\MD$ would not be able to represent each of these input patterns by a separate neuron. But nevertheless a complete and noise robust modular code emerged in $\MD$. 

A typical spike input stream from the 64 input neurons is shown in the middle row of Figure \ref{fig:fig_bars}A. The 6 squares at the bottom of Figure \ref{fig:fig_bars}A show for 6 representative time points (indicated by grey vertical lines) the resulting pixel-wise code that represents the network input (darkness of red color indicates the output trace of that input neuron at that time, that is, its output spike train convolved with the synaptic response kernel). After representing such a continuously varying input stream for 400 s to the network, subpopulations of a few neurons ($19.4$ on average) emerged that each indicated through their firing the presence of a bar at a particular position in a noise robust manner through the firing of several neurons (bar position indicated at the left side of Figure \ref{fig:fig_bars}B, and through a corresponding shading in the background of the spike raster). In this way each composite input image is represented through an emergent sparse modular neural code. This holds in spite of the fact that the image presentations were not synchronized, i.e., individual bars appeared and disappeared at random time points, and the number of simultaneously present bars varied.  

We quantified the learning performance of our model $\MD$ in extended simulations where the network was exposed to this input for 1000 s of simulated biological time. The evaluation based on 10 runs with independently drawn initial synaptic weight settings and input patterns is shown in Figure \ref{fig:fig_performance} (see {\em Methods} for details). 
Figure \ref{fig:fig_performance}A shows the number of neurons recruited for modular neural coding during learning. 
Figure \ref{fig:fig_performance}B shows that the network rapidly and robustly learns to represent all 16 bar positions.
In Figure \ref{fig:fig_performance}C, network coding performance is plotted against learning time in terms of the F1-measure \cite{van2004geometry}.
This measure is suitable for analyzing the reliability of assembly codes, where several neurons in an assembly can become selective for the same feature (here: bar position) in the network input. 
The F1-measure was separately computed for each bar position.  An F1 measure of $1$ for a bar position indicates that the bar is correctly reported by those neurons that are selective for a bar at this position, i.e., at least one of the neurons
in the corresponding emergent assembly 
is active if this bar is present and all are inactive otherwise. Hence, a high F1 measure indicates a robust encoding of bar positions by excitatory neurons in $\MD$. In Figure \ref{fig:fig_performance}C, the average F1 measure over all bar positions is plotted. After $1000$ s of learning, an average F1 measure of $0.87$ was attained. For comparison, a WTA network with idealized strong inhibition (as used for Figure \ref{fig:fig_tilted}H) was trained on the same input. In the WTA circuit, neurons did not develop a modular code but specialized on combinations of bars.
Note that the network already represents the input very well after about $200$ s of learning (see  Figure \ref{fig:fig_performance}C), although only around $200$ neurons have become pattern selective at this point (see  Figure \ref{fig:fig_performance}A). Subsequentely, the ensembles that represent basic patterns become larger, but this has only a small impact on network performance.

\begin{figure}[!t]
 \includegraphics[width=\textwidth]{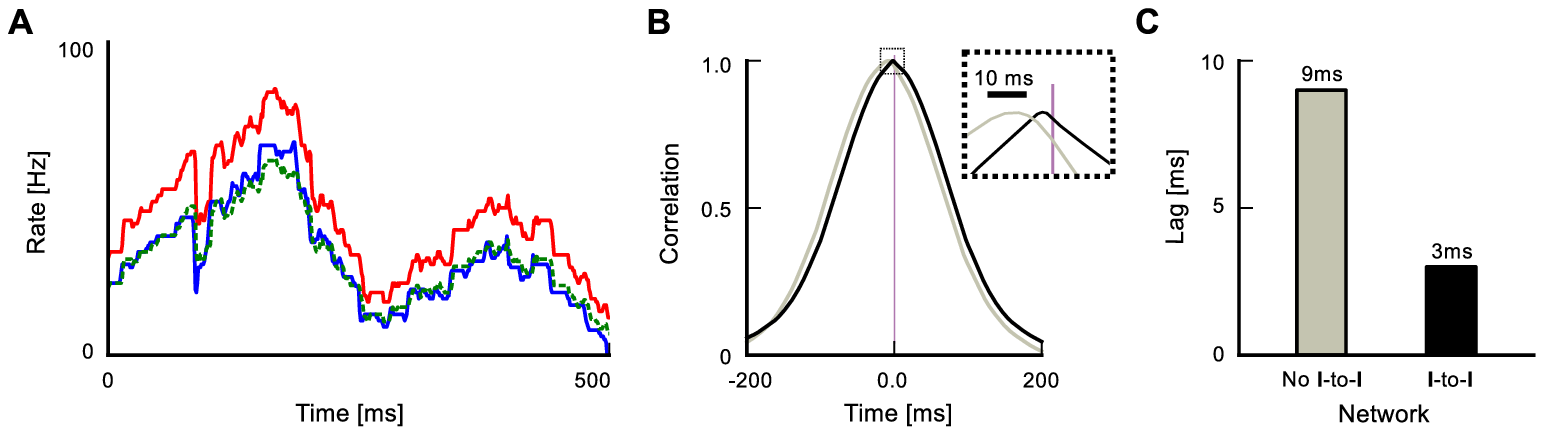}
\caption{{\bf Time course of excitation and inhibition in the model $\MD$} {\bf A}) Time course of the average firing rate of excitatory (blue) and inhibitory neurons (red) during $500$ ms after learning in the experiment shown in Figure \ref{fig:fig_stp}. The dashed green line shows a scaled version of the average inhibitory rate for better comparison. {\bf B}) 
Cross-correlation function between the excitatory and inhibitory rate reveals a small lag of about 3 ms between excitation and inhibition, comparable to in-vivo data. Shown is the cross correlation with intact connections among inhibitory neurons (black) and without these connections (gray). Inset shows a zoom into the dotted rectangle. {\bf C}) Quantification of the lag between excitation and inhibition from panel B. Intact inhibition among inhibitory neurons significantly reduces the lag between excitation and inhibition.
}
\label{fig:fig_lag}
\end{figure}

\subsection{Comparing the resulting temporal dynamics of inhibition with experimental data}

We analyzed the resulting temporal dynamics of inhibition in the model $\MD$ after 
the learning experiment of Section \ref{sec:superpositions} (Figure \ref{fig:fig_stp}). Figure \ref{fig:fig_lag}A shows the time courses of the average firing rates of excitatory neurons (blue) and inhibitory neurons (red; scaled inhibitory rate in green for better comparison) during an example time interval of $500$ msec. Consistent with experimental findings \cite{OkunLampl:08}, inhibition tracks excitation quite precisely, with a small time lag. 

We quantified the lag between excitation and inhibition like in \cite{OkunLampl:08} as the temporal offset of the peak of
the cross-correlation function between the excitatory and scaled inhibitory firing rates (plotted in Figure \ref{fig:fig_lag}B, black line). The resulting lag of $3$ msec is comparable to the measured mean lag of 3.5 ms in-vivo \cite{OkunLampl:08}.

In accordance with the data in \cite{avermann2012microcircuits} we included inhibitory connections within the pool of inhibitory neurons (I-I connections) in the microcircuit motif model $\MD$. 
We found that these connections play an important role, because they decrease the lag between excitation and inhibition. This is quantified in Figure \ref{fig:fig_lag}B,
where the black line shows the resulting correlation between excitation and inhibition in the model, and the grey line for a variation of the model where all I-I connections have been deleted.
The average lag between excitation and inhibition
increased through this deletion from 3 ms (black bar in Figure \ref{fig:fig_stp}C) to $9$ ms (gray bar in Figure \ref{fig:fig_stp}C). With intact I-I connections, inhibition is sharpened since early inhibitory responses to excitation reduce subsequent inhibitory spikes with a larger lag.
For further details on these experiments see {\em Details to simulations for Figure \ref{fig:fig_lag}} in {\em Methods}.

\section{Discussion}

We have investigated the computational properties of interconnected populations of pyramidal cells and $\text{PV}^+$ interneurons in layer 2/3 (Figure \ref{fig:fig_model}), one of the most prominent motifs in cortical neural networks. Our analysis was based on data from the Petersen Lab for layer 2/3 of mouse barrel cortex as summarized in \cite{avermann2012microcircuits}.
We have shown that the dynamics of inhibition in a simple model $\MD$ for this microcircuit motif is consistent with additional experimental data. Figure \ref{fig:fig_model}D shows that the resulting feedback inhibition is consistent with data from \cite{WilsonETAL:12}. Furthermore inhibition follows excitation in our model with a lag of around $3$ ms (see Figure \ref{fig:fig_lag}A), a value that is close to the experimentally measured mean lag of $3.5$ ms \cite{OkunLampl:08}. 
The model $\MD$ 
has produced in addition in Figure \ref{fig:fig_lag}B,C a hypothesis for the functional role of synaptic interconnections among $\text{PV}^+$ cells in this context: 
It suggests that these connections contribute to the small value of this lag.

We found that the role of inhibition in this microcircuit motif cannot be captured adequately by a WTA model.
We are proposing to consider instead a variation of the k-WTA model, where the $k$ most excited neurons are allowed to fire. The k-WTA model is well known in computational complexity theory, and tends to produce more computational power than the simple WTA model \cite{Maass:00}. A closer look shows that the dynamics of the microcircuit motif can even better be captured by an adaptive k-WTA model. In this model, the actual number of neurons that fire in response to a network input may vary. 

We have investigated the computational properties that emerge in the model $\MD$ under STDP for spike input streams that contain superimposed firing patterns. We found the emergent capability to disentangle these patterns, and represent the occurrence of each pattern by a separate sparse assembly of neurons (Figure \ref{fig:fig_tilted}-\ref{fig:fig_bars}). Hence we propose that the ubiquitous microcircuit motif of densely interconnected populations of excitatory and inhibitory neurons provides 
an important atomic computational
operation to large-scale distributed brain computations.  
Through this operation, each network module may produce one of a small repertoire of 
stereotypical firing patterns, commonly referred to as assemblies, assembly sequences, or packets of information \cite{LuczakETAL:15}.
If these assembly activations 
are fundamental tokens of global cortical computation and communication, as proposed by \cite{LuczakETAL:15}, then cortical columns have to solve a particular instance of the well-known cocktail party problem \cite{Cherry:53}: They have to recognize and 
separately represent spike inputs from different assemblies that are superimposed in their network input stream.

The existence of blind source separation mechanisms of this type had already been postulated in \cite{foldiak1990forming} as a prerequisite for avoiding a combinatorial explosion in the number of neurons that are needed to represent the information contained in complex spike input streams.
We have shown in Figs.~\ref{fig:fig_stp}--\ref{fig:fig_performance} that  
blind source separation for spike patterns emerges automatically in
the microcircuit motif model $\MD$ through STDP. This holds even for a more demanding version of the
benchmark task that \cite{foldiak1990forming} had proposed: Disentangling and representing superpositions of bars not only for a fixed number, but also for varying numbers of superimposed bars.

\subsubsection*{Relation to theoretical models for cortical microcircuit motifs}\label{sec:theory}

It is natural to ask whether a theoretical analysis can be performed to better understand the emergence of this fundamental computational capability. Unfortunately, the analysis from \cite{nessler2013bayesian} and \cite{habenschuss2013emergence} in terms of mixture distributions is only applicable to WTA circuits. A new probabilistic model  
for softer divisive inhibition 
is introduced in \cite{JonkeETAL:17}. 
The theoretical analysis in \cite{JonkeETAL:17} shows in particular that one can relate 
some parameters of the model $\MD$ --- such as the neural excitability $\alpha$ and various synaptic efficacies in the network --- to parameters of this probabilistic model. 
We used the network parameters that were derived in \cite{JonkeETAL:17} in all our simulations.

\subsubsection*{Related work}
Learning in networks of excitatory and inhibitory neurons was also studied in \cite{litwin2014formation}. They did however not study plasticity of synaptic connections from inputs to the network. Consequently, their model could not learn to perform any feature extraction from input patterns, which is the primary emergent computational property of the model $\MD$. Rather, self-organization led in the model of \cite{litwin2014formation} to an associative memory-like network behavior. An interesting feature of their model was the use of a fast Hebbian STDP rule for synaptic connections from inhibitory to excitatory neurons (iSTDP), which was in their model essential for maintaining a balance of excitation and inhibition. We did not find a need for such fast inhibitory plasticity. Instead, we set the strengths of inhibitory connections to fixed values. 
However, it would be interesting to study which types of iSTDP would lead to a self-organization of inhibitory dynamics that also supports blind source separation. 

A soft WTA model for cortical circuits with lateral inhibition was previously studied in \cite{deAlmeidaETAL:09}. Consistent with our model, the authors arrived at the conclusion that lateral inhibition in cortical circuits gives rise to an adaptive k-WTA mechanism, rather than a strict k-WTA computation. However, since it was essential for their study that the circuit operates in the limit of no noise, their model is hard to compare to the stochastic model that we have examined. Further, the authors did not incorporate synaptic plasticity into their model, which is the focus of this paper.

The model $\MD$ is also somewhat similar to the models of \cite{nessler2013bayesian, HabenschussETAL:13, kappel2014stdp}. However, these studies did not model inhibition 
through feedback from inhibitory neurons. Instead inhibition was provided in a symbolic manner as a normalization of network activity, leading to strict WTA behavior. 

The emergent computational operation in our model, the extraction of superimposed components of input patterns, is closely related to independent component analysis (ICA) \cite{HyvaerinenETAL:04}. Previous work in this direction includes the classical work by F{\"o}ldiak \cite{foldiak1990forming} and implementations of ICA in artificial neural networks \cite{hyvarinen1999fast}. It was shown in \cite{BellSejnowski:97} that ICA predicts features of neural tuning in primary visual cortex. A more recent model for a similar computational goal was proposed in \cite{lucke2010expectation}. This model is more abstract and only loosely connected to cortical microcircuit motifs.
ICA with spiking neurons was previously considered in \cite{savin2010independent}. The authors derived theoretical rules for intrinsic plasticity (i.e., rules for homeostasis of neurons) which, when combined with input normalization, weight scaling, and STDP, enable each neuron to extract one of a set of independent components of inputs. While closely related in terms of the computational function, the 
data-based form of inhibition in our model $\MD$ has quite different features. In \cite{savin2010independent}, the main purpose of inhibition is to decorrelate neuronal activity so that different neurons extract different features. Sparse activity is enforced there by intrinsic plasticity. Intrinsic plasticity in their model is thus required to work on a fast time-scale (the time scale of input presentations). In contrast, sparse network activity in our data-based model $\MD$ is enforced by inhibition. 
It is known that feedback inhibition is very fast and precise \cite{OkunLampl:08}, while it is unclear whether this is also true for intrinsic plasticity \cite{turrigiano2004homeostatic}. 

\subsubsection*{Experimentally testable predictions of our model}

A main prediction of our model (see Figure \ref{fig:fig_stp}) is the emergence of blind source separation of superimposed spike patterns.
In addition, our model predicts that each of the identified basic patters of the spike inputs becomes represented through some separate assembly of pyramidal cells. Our model predicts that this effect takes place for any type of network input, e.g. also for artificially generated stimuli that the organism is never exposed to in a natural environment. This hypothesis can be tested experimentally, e.g. through optogenetic control. 

In addition our model predicts a specific role of synaptic connections among PV$^+$ inhibitory cells (see Figure \ref{fig:fig_lag}): They contribute to the experimentally found small time lag of just a few ms by which inhibition trails excitation. This 
prediction can be tested experimentally by silencing synaptic connections among PV$^+$ cells and measuring the impact on the lag between excitation and inhibition.

\vspace{3cm}

\subsubsection*{Acknowledgments}

Written under partial support by the Human Brain Project of the European Union \#604102 and \#720270. We would like to thank Carl Petersen for helpful discussions. 

\clearpage

\bibliographystyle{apalike}

\end{document}